\documentclass[11pt]{article} 

\usepackage{amsmath,amsthm,latexsym,amssymb,amsfonts,epsfig}

\usepackage{float}
\usepackage{upgreek}
\usepackage{cite}

\usepackage{xcolor}
\usepackage{amsmath}
\usepackage{amsthm}
\usepackage{graphicx}
\usepackage{amssymb}
\usepackage{physics}
\usepackage{tensor}
\usepackage{mathtools}
\usepackage{dsfont}
\usepackage{color}
\usepackage{soul}
\usepackage[hyperfootnotes=true,unicode=true, 
 bookmarks=true,bookmarksnumbered=false,bookmarksopen=false,
 breaklinks=false,pdfborder={0 0 1},backref=false,colorlinks=true]
 {hyperref}
 \usepackage{float}
 \usepackage[markup=nocolor]{changes}

\usepackage[makeroom]{cancel}

\usepackage{etoolbox}

\usepackage{amsfonts}

\newcommand{\overbar}[1]{\mkern 1.5mu\overline{\mkern-7mu#1\mkern-1.5mu}\mkern 1.5mu}

\usepackage{hyperref}
\hypersetup{
    colorlinks,%
    citecolor=blue,%
    filecolor=blue,%
    linkcolor=blue,%
    urlcolor=blue
}

\newcommand{\m}{\mbox{}}

\newcommand{\be}{\begin{equation}}
\newcommand{\ee}{\end{equation}}
\newcommand{\ba}{\begin{eqnarray}}
\newcommand{\ea}{\end{eqnarray}}
\def\bal#1\eal{\begin{align}#1\end{align}}

\makeatletter
\@addtoreset{equation}{section}
\makeatletter
\@ifundefined{textcolor}{}
{%
 \definecolor{BLACK}{gray}{0}
 \definecolor{WHITE}{gray}{1}
 \definecolor{RED}{rgb}{1,0,0}
 \definecolor{GREEN}{RGB}{0,204,0}
 \definecolor{BLUE}{rgb}{0,0,1}
 \definecolor{CYAN}{cmyk}{1,0,0,0}
 \definecolor{MAGENTA}{cmyk}{0,1,0,0}
 \definecolor{YELLOW}{cmyk}{0,0,1,0}
 }

\usepackage{hyperref}
\hypersetup{
    colorlinks,%
    citecolor=blue,%
    filecolor=blue,%
    linkcolor=blue,%
    urlcolor=blue
}

\makeatother

\newcommand{\re}{\mathbb{R}}

\NewDocumentCommand\func{moo}{\opbraces{\mathop{#1\m}\nolimits\IfNoValueTF{#2}{}{^{#2}}\IfNoValueTF{#3}{}{_{#3}}}}
\NewDocumentCommand\fomega{o}{\func{\omega}[][#1]}
\NewDocumentCommand\fOmega{o}{\func{\Omega}[#1]}

\begin{document}

\title{ Effective LQC model for $k=+1$ isotropic cosmologies from spatial discretisations}

\author{Klaus Liegener$^{1,2}$\thanks{klaus.liegener@desy.de} , Stefan Andreas Weigl$^3$\thanks{stefan.weigl@fau.de}
\\
\\
{ $^1$ Dept. of Physics and Astronomy, Louisiana State University}\\
{ 70808 Baton Rouge,  USA}\\
{$^2$ II. Institute for Theoretical Physics, University of Hamburg,}\\
{Luruper Chausee 149, 22761 Hamburg, Germany}\\
{$^3$ Institute for Quantum Gravity, FAU Erlangen-N\"urnberg,}\\
{Staudtstr. 7, 91058 Erlangen, Germany}\\
}



\date{\today{}}

\maketitle
\begin{abstract}

\fontfamily{lmss}\selectfont{
The closed, spatially isotropic FLRW universe ($k=+1$) is endowed with modifications due to a discrete underlying space-structure. Motivated from Loop Quantum Gravity techniques, a full Thiemann regularisation is performed. The impact of these modifications of the single-graph-sector appearing in the scalar constraint are interpreted as physical quantum gravity effects. We investigate the form of the modified scalar constraint and its analytical approximations for $k=+1$ spacetimes and assume this effective constraint as the generator of dynamics on the reduced isotropic phase space. It transpires that the system still features a classical recollapse with only marginal discreteness corrections. Moreover, the initial and final singularities are resolved and we present an effective model mirroring the qualitative features of system.
}
\end{abstract}

\section{Introduction}
\label{s1}
\numberwithin{equation}{section}
A prominent example for solutions of classical general relativity (GR) are the cosmological spacetimes of Friedmann-Lema\^{i}tre-Robertson-Walker (FLRW) type \cite{Fr22,Fr23,Fr24,Le27,Le31,Le33,Rob35,Rob36a,Rob36b,Wal37}. This is on the one hand due to their precise agreement with observations \cite{Muk05,dS34} and on the other hand because  they allow us to reduce the field-content of GR to finitely many degrees of freedom due to their symmetry. Consequently, they are also often the focus of candidate theories of quantum gravity: while a complete quantisation of GR is not available as of today, it is hoped that for these special cases at least partial results can be obtained. Following this spirit, many works concern the isotropic sector of Loop Quantum Gravity (LQG) \cite{GP00,Rov04,AL04,Thi09}, a canonical quantisation of GR expressed in its $SU(2)$-connection formulation \cite{Ash86,Ash87,Bar94,Bar95}. Especially for the case $k=0$, i.e. spatially flat isotropic spacetimes, a big subfield emerged -- called Loop Quantum Cosmology (LQC) \cite{Boj99a,Boj99b,ABL03,Boj05,Ash08} -- where techniques similar to LQG are used to directly quantise the symmetry-reduced phase space of FLRW. Yet its relation to full LQG is as of today unclear but remains an active field of research \cite{Engle07,Fle15,BEHM16}.\\
However, due to the connection formulation of GR on which LQG is based, one has many similarities to the framework of lattice gauge theories (LGT) and could look at discretisations of space similar to the Hamiltonian language of LGT. In the presence of a finite ultraviolet cutoff, the scalar (or Hamiltonian) constraint of GR has to be discretised, i.e. approximated by a function expressed solely in terms of quantities on the lattice such as holonomies and fluxes. While many such discretisations are in general possible, the first one being suitable for quantisation came from Thiemann in his seminal papers \cite{Thi98a,Thi98b}. Such a discretisation can be recovered if one takes the expectation value of the scalar constraint operator in coherent states supported on a lattice of the kinematical LQG Hilbert space \cite{AQG1,AQG2}. Said procedure has shown itself to be especially of interest, if one takes the coherent states to be peaked over flat, isotropic cosmology \cite{AC13,AC14a,AC14b,DL17a,DL17b}: in the case of Thiemann-regularisation one obtained a function in terms of the parameters of the FLRW-reduced phase space that resembles an effective scalar constraint  which describes the resolution of the initial singularity in terms of a big bounce on the LQC Hilbert space \cite{ADLP18,ADLP19} (see \cite{GM19a,GM19b} for extensions of this to Bianchi I spacetimes). The resolutions of singularities via evolution operators in the reduced theory of LQC had already a long history \cite{APS06a,APS06c,Boj08}, however it used to focus in earlier works on a quantum operator whose discretisation did {\it not} stem from the full theory, but rather first imposed symmetries of $k=0$ spacetimes prior to discretisation. Afterwards, the resulting dynamics differed from the one, one would obtain if one first discretises the theory and imposes cosmology a posteriori: most notably the ``first-reduce-then-regularising'' framework lead to a symmetric bounce, while the bounce was genuinely asymmetric for the ``first-regularise-then-discretise'' framework. But since only the later one agrees with the expectation value of the Thiemann-regularisation in suitable coherent states for flat isotropic cosmology, the question arises how the situation behaves for models involving nontrivial spatial curvature\footnote{We like to point out that a second regularisation of the scalar-constraint operator exists in the literature \cite{Warsaw1,Warsaw2} which appears to work well for spatially flat spacetimes and is expected to reproduce the original LQC effective scalar constraint as expectation values. However, due to limited studies of its semi-classical properties for spatially curved spacetimes, we refrain from using it for the purpose of this article.}. In this paper we will extend therefore the analysis to closed isotropic spacetimes, characterised by $k=+1$.\\ 

Few works in the literature have dealt so far with the $k=+1$ sector and those either proposed modifications of the reduced constraint ad hoc \cite{ST04} or considered LQC-like quantisations \cite{SKL06,APSV07,MHS09}. In the later approach, one again followed -- among other simplifications -- the philosophy of implementing symmetries of isotropic curved spacetimes prior to the discretisation process. In this paper we will refrain from these simplifications and rather ask about the properties of an effective scalar constraint in line with the regularisation techniques of Thiemann. For this purpose, we will stay purely classical and base the relation to LQG merely on the kinematical expectation value of the scalar constraint operator taken in the complexifier-coherent states from Thiemann \& Winkler \cite{TW1,TW2,TW3}. According to several studies \cite{AQG2,DL17b} in leading order of $\hbar$ this expectation value will agree with its classical discretisation thereby justifying our classical computations. Due to the complexity of the framework involved, we are forced to truncate this expectation value at finite orders in the lattice spacing in order to obtain analytical closed formulas.\\
As of today, the full quantum dynamics is computationally out of reach, but to gain a first intuition of the dynamics (up to higher $\hbar$-corrections) we utilise the {\it effective} LQC programme. Here, the regularised functions are taken as effective scalar constraints on some reduced phase space characterised by the cosmological parameters \footnote{There is exhaustive literature concerning justification of such procedures, see e.g. \cite{Tav08,BS06,GP19}.}. That means, that the modifications due to the discrete nature of spacetime are interpreted as {\it physical} predictions of a theory of quantum gravity! For the purpose of comparison, we will adopt this point of view inside a toy model as well. To do this, we insert an additional assumption into our framework, namely that {\it the dynamics of a discretised lattice theory of finitely many degrees of freedom driven by a discretised scalar constraint for symmetric initial conditions is assumed to agree with the dynamics of the symmetry-reduced constraint on the symmetry-reduced phase space}.  Under this premise -- which is crucial for any effective model -- we compute the flow of the aforementioned truncations of the scalar constraint and study its behaviour in the classical regime. To the exact regularisation for the scalar constraint of $k=+1$ cosmology we have only access numerically, nevertheless we will use numerical investigations to judge the quality of the analytical, approximated constraints and in order to motivate an {\it effective} constraint, which captures the qualitative features of the full evolution.
\\

The organisation of this article is as follows:\\
In \autoref{s2} we repeat the formulation of GR in terms of the Ashtekar-Barbero variables as a SU(2) gauge theory. We introduce a discretisation of our manifold, in terms of a hyperspherical lattice. The fundamental building blocks of the continuum theory, i.e. the connection and its conjugate momentum, are replaced by suitable smearings on the lattice, namely holonomies and fluxes. When specialising to $k=+1$ cosmology, one can give closed analytical formulas for them along all edges of the graph. Afterwards, we will turn to $C^\epsilon$, the Thiemann-regularisation of the scalar constraint. Due to the fact that the holonomies carry an explicit coordinate dependence, evaluating $C^\epsilon$ on the hyperspherical lattice for $k=+1$ cosmology results in lengthy expressions that are not handleable analytically. Therefore, we merely present an expansion of $C^\epsilon$ up to $7 ^{\rm th}$ order in terms of the regularisation parameter $\epsilon$, i.e. the lattice spacing.\\
In \autoref{s3} we couple the regularised scalar constraint to a free, massless and homogeneous scalar field. The resulting expression is postulated to describe the dynamics of a regularised classical system initially found to be in the phase space point derived in \autoref{s2}. We will then investigate the dynamics of the numerically evaluated full theory and compare it to those of the approximated constraints. In this model, the discretisation itself resolves the initial and final singularities of the universe and instead joins it with forever expanding/contracting universe via so-called big bounces. This behaviour can be encapsulated in a comparable simple {\it effective} model for which we display the analytic formulas.\\
In \autoref{s4} we conclude and discuss future research directions.\\

\section{Connection formulation of general relativity and its discretisation}
\label{s2}
\numberwithin{equation}{section}
In the first subsection, the Ashtekar-Barbero formulation of GR \cite{Ash86,Ash87,Bar94,Bar95} is reviewed and discretised on a purely classical level. For details we refer to the literature, e.g. \cite{Thi09}. Afterwards, we will investigate a possible incarnation for such a discretisation explicitly in the context of closed, isotropic spacetimes.\\

\subsection{Ashtekar-Barbero formulation on a fixed lattice}
One of the major steps towards defining a theory of quantum gravity, was the realisation that the Hamiltonian formulation of GR can be understood as an SU(2) gauge theory of Yang-Mills type:\\
Let $\sigma$ be a 3-dimensional, spatial, orientable, compact\footnote{Compactness is not a requirement for the general framework to work. However, as we are interested in closed $k=+1$ models in the present paper, we will restrict to compact manifold from the onset.} manifold.
The Ashtekar-Barbero variables coordinatise the phase space $\mathcal{M}$ of an SU(2) Yang-Mills theory, described by an SU(2) connection (gauge potential) $A_a (x) := A^J_a(x)\tau_J\; :\; \sigma \to \mathfrak{su}(2)$ and a non-Abelian conjugate momenta $E^a(x):= E^a_J(x)\tau_J\;:\; \sigma \to\mathfrak{su}(2)$, for which we choose positive orientation (with spatial indices $a,b,..=1,2,3$ and internal $\mathfrak{su}(2)$ indices $I,J,...=1,2,3$ and $\tau_J$ being $-i/2$ times the Pauli matrices). The symplectic form on $\mathcal{M}$ is given by
\begin{align}
\omega =\frac{2}{\kappa\beta}\int_\sigma {\rm d}^3x\; dE^a_I(x)\wedge dA^I_a(x)
\end{align}
which leads to the Poisson algebra:
\begin{align}\label{poissoinalgebraashtekar}
\{ E^a_J (x), E^b_K(y)\} =\{A^J_a(x),A^K_b(y)\}=0,\hspace{30pt} \{E^a_K(x), A^J_b(y)\}=\frac{\kappa\beta}{2}\delta^a_b\delta^J_K\delta^{(3)}(x,y)
\end{align}
with $\kappa=16\pi G$ and $\beta\in\mathbb{R}-\{0\}$, the so-called Immirzi parameter. This phase space is subject to the Gauss constraint:
\begin{align}
G_J=\mathcal D_a E^a_J =\partial_a E^a_J +\epsilon_{JKL} A^K_a E^a_L =0.
\end{align}
The phase space of Ashtekar-Barbero variables becomes equivalent with the phase space of the Hamiltonian framework of GR as long as the Gauss constraint is satisfied.\\
The other constraints of GR read in this framework:
\begin{itemize}
\item The diffeomorphism (or vector) constraint:
\begin{align}
D_a=\frac{2}{\kappa\beta}F^J_{ab}E^b_J+D_{a, {\rm matter}}
\end{align}
\item The scalar (or Hamiltonian) constraint
\begin{align}\label{scalar_constraint}
C=\frac{1}{\kappa}\left(
F^J_{ab}-(1+\beta^2)K_{ac}e^c_MK_{bd}e^d_N\epsilon_{MNJ}
\right)\epsilon_{JKL}\frac{E^a_K E^b_L}{\sqrt{\det(E)}}+C_{\rm matter}
\end{align}
\end{itemize}
with $F_{ab}$ the curvature of the Ashtekar connection and $K_{ab}$ the extrinsic curvature and the spin connection $\Gamma^J_a$:
\begin{align}\label{Definitionextrinsiccurvature}
F_{ab}^J&:=2\partial_{[a}A^J_{b]}+\epsilon^J_{\, KL} A^K_aA^L_b,\hspace{30pt}
K_{ab}:=\beta^{-1 }e^J_b(A^J_a-\Gamma^J_a(E))
\end{align}
The terms $D_{\rm matter}$ and $C_{\rm matter}$ correspond to the applicable matter content. For the purpose of this section we will set them to zero and discretise only vacuum GR. These quantities will reappear in \autoref{s3} where the effective dynamics of a concrete cosmological model is studied in the presence of a free massless scalar field.\\

Before we turn towards discretisation, a crucial step is the possibility to rewrite the scalar constraint (\ref{scalar_constraint}) in such a way that the inverse power of $\det(E)$ is taken care of. This has important advantages if one goes to the quantisation of the theory and therefore serves as the starting point of modern dynamics in LQG. This transformation is achieved via the first of the famous Thiemann identities \cite{Thi98a,Thi98b}:
\begin{align}
\{V, A^J_a\} =\frac{\kappa \beta}{8}\epsilon^{JKL}\epsilon_{abc}\frac{E^b_KE^c_L}{\sqrt{\det(E)}}\\
\{\{V,C_E[1]\},A^J_a\}=\frac{\kappa\beta^3}{2}K_{ab}e^b_J
\end{align}
where $V=\int_\sigma \sqrt{\det(E)}$ is the volume of the spatial manifold.\\
Note that the second identity helps to express the function $K_{ab}$ which is originally a complicated object in terms of the connection $A$, the momentum $E$ and its derivatives, as an expression that does not depend on the derivatives anymore.\\

To stay maximally close the framework of Lattice Gauge Theories the basic variables, to start the quantisation procedure with, is not the connection, but rather its smearing along edges of some graph. Its conjugate momentum will be smeared along surfaces of the dual cell complex, respectively.\\
Let $\gamma\subset \sigma$ be a graph, that is a collection of edges $e : [0,1] \to \sigma$ meeting at most at their endpoints, and such that $\gamma$ allows for a dual cell complex, i.e. one can associate to each vertex $\nu$ three faces $S_e$ such that $S_e\cap e'= \nu$ and normal to $\dot{e}'$ if $e'$ is in direction $e$. For each such $\gamma$ we will now introduce a phase space by considering the collection of discretised phase space variables associated to each edge of $\gamma$ following the constructions of Lattice Gauge Theory:\\
Along the edges $e$ of the lattice, we will compute the holonomies $h(e)\in {\rm SU}(2)$ of the connection  i.e. the path ordered exponential
\begin{align}
h(e):= \mathcal{P}\exp \left( \int_e {\rm d}x^a\; A^J_a(x)\tau_J \right)\,
\end{align}
where later values are ordered to the right, and the fluxes along the associated surfaces $S_e$:\footnote{A different construction would be gauge-covariant fluxes. Instead of the smearing used here, one would consider one which transforms feasibly under gauge transformations even for finite graphs. However, for the purpose of this paper we will stick to regular fluxes and refer to \cite{Thi00} for their possible implications.}
\begin{align}
E(e)= E_I(e)\tau_I :=\int_{S_e} {\rm d}x^a\wedge {\rm d}x^b \epsilon_{abc} E^c_I(x)\tau_I
\end{align}
For the purpose of further studies we will consider in this paper families of lattices $\{\gamma_\epsilon\}_{\epsilon\in\mathbb{R}}$ parametrised by their lattice spacing $\epsilon$. The $\gamma_\epsilon$ are of the form that they are (i) cuboidal (i.e. at each vertex 6 edges meet), (ii) can form directed families of subsets amongst each other $\gamma_{2\epsilon}\subset \gamma_{\epsilon}$  and (iii) lie dense in $\sigma$, in the sense that each open neighbourhood will be punctured by $\gamma_{\epsilon}$ for $\epsilon$ small enough.\\

We will now discretise the scalar constraint, i.e. we will search for a function $C^\epsilon$ that is completely expressed in terms of the discretised phase space variables of graph $\gamma_\epsilon$ and such that in the limit $\epsilon\to 0$ the original scalar constraint is restored. To be precise, in terms of a smearing against a function $N$ of compact support, we want:
\begin{align}
\lim_{\epsilon\to 0} \sum_{v\in\gamma_\epsilon} N(v)\; C^\epsilon(v) = \int_\sigma {\rm d}x^3 \; N(x)\; C(x) 
\end{align}
Of course, there are several possibilities for $C^\epsilon$ and the discretisation we will study in this paper is the graph-preserving version \cite{AQG1} of the original regularisation of Thiemann \cite{Thi98a,Thi98b}:
\begin{align}\label{Lorentzian}
C^\epsilon[N] &= C^\epsilon_E[N]+\frac{2^3(1+\beta^2)}{\kappa^4\beta^7}\sum_v N(v)\sum_{ijk}\epsilon(i,j,k) \times\nonumber\\
&\;\;\;\times \Tr\bigg[
h(e_i)\Big\{h(e_i)^\dagger, \{V,C^\epsilon_E[1]\}\Big\}h(e_j)\Big\{h^\dagger(e_j),\{V,C^\epsilon_E[1]\}\Big\}h(e_k)\Big\{h^\dagger(e_k),V\Big\}
\bigg],\\
C^\epsilon_E[N] &=\frac{-1}{2\kappa^2\beta}\sum_{v}N(V)\sum_{ijk}\epsilon(i,j,k)\Tr\bigg[
\left(h(\Box^\epsilon_{v,ij})-h^\dagger(\Box^\epsilon_{v,ij})\right)h(e_k)\Big\{h^\dagger(e_k),V\Big\}
\bigg]\label{Euclidian}
\end{align}
with $\epsilon(a,b,c) :=\mathrm{sign}\Big( \det(\dot{e}^a(0),\dot{e}^b(0),\dot{e}^c(0))\Big)= \mathrm{sgn}(a)\,\mathrm{sgn}(b)\,\mathrm{sgn}(c) \,\tensor{\epsilon}{^{\qty|a|}^{\qty|b|}^{\qty|c|}}$ being a generalised epsilon tensor, which sums over negative indices respecting their sign, too.
A similar construction can be carried out for functions whose vanishing is equivalent to the vanishing of the diffeomorphism constraint \cite{AQG1}: Instead of discretising $D_a$ itself, one can consider $\tilde{D}_I:= E^a_I D_a$ which vanishes iff the vector constraint vanishes due to the nondegeneracy of $E^a_I$. The form of $\tilde{D}_I$ is more suited for quantisation due to the fact that its discretisation has no explicit dependence of the regulator. However, since we are interested in isotropic cosmology in the following, and the diffeomorphism constraint vanishes trivially there, we will refrain from considering these expressions explicitly.\\
The regularised scalar constraint $C^\epsilon$ is an approximation to the generator of time-gauge-translations in the continuum and we will promote it to the generator of time-translations on $T^\star \mathcal M (\gamma_\epsilon)$, the discretised phase space of the graph.\\
The Poisson algebra of the holonomies and fluxes is given by:
\begin{align}\label{holonomzfluxalgebra}
  \big\{ E_I(e), E_J(\tilde{e})\big\} =\big\{h(e), h(\tilde{e})\big\}=0 \hspace{30pt}\big\{E_I(\tilde{e}), h(e)\big\} = \frac{\beta \kappa}{4}\Big( \sigma(e(0),S_{\tilde{e}}) \tensor{\tau}{_I} h(\tensor{e}{})  +  \sigma(e(1),S_{\tilde{e}})h(\tensor{e}{})\tensor{\tau}{_I} \Big),
\end{align}
where $\sigma(x,S)$ is one if $x\cap S\neq \emptyset$, otherwise vanishing. 
Note that the Poisson brackets of these smeared variables have lost the distributional character of \eqref{poissoinalgebraashtekar} and have thus a much more suitable form for a quantization.

\subsection{The case of isotropic, closed cosmologies}
This subsection we apply the above developed framework explicitly to the case of compact spacetimes that are spatially isotropic and homogenous. In order to allow for an isotropic metric, the spatial manifold needs to be of the form $\sigma \cong S^3$.\\
In terms of canonical phase space variables, the spatial metric can be written as the conformal line element:
\begin{align}
q= q_{ab} {\rm d} x^a{\rm d}x^b =\frac{a(t)^2\delta_{ab}}{\left(1+\frac{1}{4}\tilde{r}^2\right)^2}{\rm d}x^a {\rm d}x^b
\end{align}
with $\tilde{r}^2=\sum_i(x^i)^2$ and $a(t)$ being the scale factor. Upon a change to another set of  coordinates
\begin{align}
\dfrac{\tilde{r}}{1+\dfrac{1}{4}\tilde{r}^2}= \sin (r),\hspace{30pt}{\rm with}\;\; r\in \big[0,\tfrac{\pi}{2}\big)\;\; ,
\end{align}
which from now on we will refer to as hyperspherical ones, we receive: ($\theta\in[0,\pi],\varphi\in[0,2\pi]$)
\begin{align}
q = a(t)^2 \left(
dr^2+\sin[2](r)({\rm d}\theta^2+\sin[2](\theta){\rm d}\varphi^2)
\right)
\end{align}
Since the space is closed, we can compute the associated finite volume of it:
\begin{align}
V=\int_0^{\tfrac{\pi}{2}}{\rm d}r \int_0^\pi{\rm d} \theta \int_0^{2\pi}{\rm d}\varphi \sqrt{\det(q)}=\pi^2 a^3
\end{align}
Starting from this line element, one can immediately compute the connection and its momentum of the Ashtekar-Barbero variables. First, we note that a possible choice of triads (defined by the relation $q_{ab}=e^J_a\delta_{JI}e^I_b$) are
\begin{align}
e_1 =\frac{\rm{d}r}{a(t)},\hspace{40pt}e_2 =\frac{{\rm d}\theta}{\sin(r)a(t)},\hspace{40pt}e_3 =\frac{{\rm d}\varphi}{\sin(r)\sin(\theta)a(t)}
\end{align}
Computing the spin connection from the triads, we obtain the following form:
\begin{align}
\Gamma^L_a&=-\frac{1}{2}\epsilon^{LJK}e^b_K(\partial_b e^J_a -\partial_a e^J_b+e^c_Je^M_a \partial_b e^M_c)\\
&= - \tensor{\epsilon}{^L^J^K}\qty(\tensor*{\delta}{^2_a}\tensor*{\delta}{_J^2}\tensor*{\delta}{^1_K}\cos(r) + \tensor*{\delta}{^3_a}\tensor*{\delta}{_J^3}\tensor*{\delta}{^1_K}\cos(r)\sin(\theta) + \tensor*{\delta}{^3_a}\tensor*{\delta}{_J^3}\tensor*{\delta}{^2_K}\cos(\theta))
\end{align}
The extrinsic curvature $\tensor{K}{_a_b}$ given by \eqref{Definitionextrinsiccurvature} is in our case quite simple. Here we do not want the action of any spatial diffeomorphisms, so we set $\tensor{N}{^a}\equiv 0$. In addition the only time dependent quantity in our metric is the scale factor $a(t)$, it is easy to check that 
\begin{align}
    \tensor{\Dot{q}}{_a_b} = 2 \frac{\Dot{a}(t)}{a(t)}\,\tensor{q}{_a_b}
\end{align} holds.\\
We will further pick a comoving frame, i.e. $N\equiv1$. With this knowledge we can calculate the connection and the electric field:
\begin{align}\label{classical_connection}
       \nonumber \tensor*{A}{^1_1} &= \beta c  & \tensor*{A}{^1_2}  &= 0  & \tensor*{A}{^1_3} &= \cos(\theta)\\
        \tensor*{A}{^2_1} &= 0  & \tensor*{A}{^2_2} &= \beta c \sin(r)  & \tensor*{A}{^2_3} &= - \cos(r) \sin(\theta)\\
     \nonumber   \tensor*{A}{^3_1} &= 0 & \tensor*{A}{^3_2} &= \cos(r)  & \tensor*{A}{^3_3} &= \beta c \sin(r) \sin(\theta)
     \end{align}\vspace{-0.5cm}
       \begin{align}\label{classical_triad}
       \tensor*{E}{^a_L}= p \tensor*{\delta}{^a_L}\bigg( \tensor*{\delta}{^a_1}\sin[2](r) \sin(\theta) +  \tensor*{\delta}{^a_2}  \sin(r) \sin(\theta) +  \tensor*{\delta}{^a_3}  \sin(r)  \bigg) .
    \end{align}
Here, we have changed the notation
\begin{align}
\dot{a}(t) \to c,\hspace{30pt} a(t)^2 \to p
\end{align}
From this point onwards, we will be interested only in families of connection and triads that are parametrized by $(c,p)\in \re^2$, i.e. of the form (\ref{classical_connection}) and (\ref{classical_triad}). These are the classical phase space data that allow an isotropic metric. However, we will not longer employ any relation to $\dot{a}(t)$, which might no longer hold in the quantum theory, where the evolution is given by the quantum Hamilton equations. In order to perform a symplectic reduction to the submanifold spanned by $(c,p)$, such an identification is not necessary.\\
As we will in the following transcend from the continuum to the discrete lattice, we recall classical, continuum formula for the scalar constraint in terms of the Ashtekar-Barbero variables of vacuum isotropic, closed cosmology:
\begin{align}
C=-\frac{6\pi^2}{\kappa}\;\sqrt{p} \;(1+c^2/\beta^2)
\end{align}

The hyperspherical lattice is composed of edges along the three coordinates lines $r,\theta,\varphi$. The degenerate points at $r=0,\tfrac{\pi}{2}$ and $\theta = 0,\pi$ are the only noncuboidal points of the lattice. Their valency is directly related to the denseness of the lattice in $\theta$- and/or $\varphi$-direction. We define the number of vertices in each direction $i$ by $N_i$. The lattice spacing $\epsilon$ has to be chosen such that
\begin{align}\label{vertices_epsilon}
\epsilon_1 =\frac{\pi}{2N_r}, \hspace{20pt} \epsilon_2 =\frac{\pi}{N_\theta},\hspace{30pt} \epsilon_3 =\frac{2\pi}{N_\varphi}
\end{align}
With these notations, one can compute for an edge $e_{i}$ which starts at $e(0)=(r_0,\theta_0,\varphi_0)$ and goes along direction $i$ for the length $\epsilon_i$:
\begin{align}
        &h(e_{\hat{r}})= \mathrm{exp}\bigg[\epsilon_1 \beta c \tau_1  \bigg]  \\
        & h(e_{\hat{\theta}}) = \mathrm{exp}\bigg[\epsilon_2 \Big(\beta c \sin(r_0) \tau_2 + \cos(r_0)\tau_3\Big)\bigg]\\
        & h(e_{\hat{\varphi}}) = \mathrm{exp}\bigg[\epsilon_3 \Big(\cos(\theta_0) \tau_1 - \sin(\theta_0)\cos( r_0) \tau_2 + \beta c \sin(r_0) \sin(\theta_0) \tau_3\Big)\bigg]
    \end{align}
The surface $S_{e_i}$ associate to any each $e_{\hat i}$ are such that the intersect the edge at its starting point and lie in the $jk$-plane normal to $i$ with boundaries $e(0)^j\pm \epsilon_j/2$ and similar for $k$. Therefore:

\begin{align}\label{FLUXES}
   E(e_r) &=\iint_{{S_{e_r}}{}} \mathrm{d}\tensor{x}{^a} \wedge \mathrm{d}\tensor{x}{^b} \, \tensor*{E}{^c_I}\tau_I\,\tensor{\epsilon}{_a_b_c} = \int\limits^{\theta_0+\frac{\epsilon_2}{2}}_{\theta_0-\frac{\epsilon_2}{2}}\mathrm{d}\theta\int\limits^{\varphi_0+\frac{\epsilon_3}{2}}_{\varphi_0-\frac{\epsilon_3}{2}}\mathrm{d}\varphi\,\sqrt{p}^3 \sin[2](r_0) \sin(\theta) \frac{1}{\sqrt{p}}\tau_1 \nonumber \\ &= p \sin[2](r_{0}) \Big(\cos(\theta_0-\tfrac{\epsilon_2}{2}) - \cos(\theta_{0}+\tfrac{\epsilon_2}{2})\Big) \epsilon_3\tau_1 \\
    E(e_\theta) &=\int\limits^{r_0+\frac{\epsilon_1}{2}}_{r_0-\frac{\epsilon_1}{2}}\mathrm{d}r\int\limits^{\varphi_0+\frac{\epsilon_3}{2}}_{\varphi_0-\frac{\epsilon_3}{2}}\mathrm{d}\varphi\,\sqrt{p}^3 \sin(r)^2 \sin(\theta_0) \frac{\tau_2}{\sqrt{p} \sin(r)} \nonumber \\ &= p \Big(\cos(r_0-\tfrac{\epsilon_1}{2}) - \cos(r_{0}+\tfrac{\epsilon_1}{2})\Big) \sin(\theta_0) \epsilon_3 \tau_2\\
    E(e_\varphi) &=\int\limits^{r_0+\frac{\epsilon_1}{2}}_{r_0-\frac{\epsilon_1}{2}}\mathrm{d}r\int\limits^{\theta_0+\frac{\epsilon_2}{2}}_{\theta_0-\frac{\epsilon_2}{2}}\mathrm{d}\theta\,\sqrt{p}^3 \sin(r)^2 \sin(\theta) \frac{\tau_3}{\sqrt{p} \sin(r) \sin{\theta}}\nonumber \\ &= p \Big(\cos(r_0-\tfrac{\epsilon_1}{2}) - \cos(r_{0}+\tfrac{\epsilon_1}{2})\Big) \epsilon_2 \tau_3
\end{align}
Having now complete knowledge of the building blocks available, one is in principle able to compute the values of discretised observables. Exemplarily, we present the computation of the discretised spatial volume , i.e.:
\begin{align}
V := \sum_{v\in\gamma} V(v),\hspace{30pt} V(v):=\sqrt{\abs{\frac{1}{3!}\sum_{e_a\cap e_b\cap e_c =v}\epsilon^{IJK} \epsilon_{abc} E_I(e_a)E_J(e_b)E_K(e_c)}}
\end{align}
This is the regularisation which leads to the Ashtekar-Lewandowski volume upon quantisation in LQG \cite{AL98}.\\
For the fluxes of the form (\ref{FLUXES}) on the hyperspherical lattice, the volume can be exactly computed as:
\begin{align}
V_0= \sqrt{p}^3 \qty(\frac{\pi}{2\epsilon_1}-1)\sin(\frac{\epsilon_1}{2}) \cot(\frac{\epsilon_2}{2}) \sqrt{2 \epsilon_2\sin(\frac{\epsilon_2}{2})} \,2 \pi
\end{align}
One can check that this indeed has the correct classical limit when removing the regulators $\epsilon\rightarrow 0$.
\\

It must be noted that it is a feature of classical, isotropic, closed cosmology, that connection and triad are {\it at all times} given by the form (\ref{classical_connection}) and (\ref{classical_triad}). Once, we go over to the discrete level, this property is not automatically guaranteed and requires further investigation \cite{DKL20,HL19}.\\
One realizes that the submanifold  $\, \overbar{\mathcal{M}}$ of the phase space spanned by connections and triads parametrised by $(p,c)$ can be understood as a symplectic reduction. That means, one can naturally endow $\, \overbar{\mathcal{M}}$ with a symplectic structure coming from $(h(e),E(e))$.\\
For the case of the continuum geometry this method of symplectic reduction can be easily applied and we repeat the computation for completeness: Let $f_1, f_2$ be functions on the phase space $\mathcal{M}$ of the continuous connection and $X_f$ the Hamiltonian flow generated by $f$, then we can derive:
\begin{align*}
    \big\{f_1, f_2\big\} &= \int_\sigma \mathrm{d}^3x\,2\fdv{f_{[1}}{\tensor*{E}{^a_I}}\fdv{f_{2]}}{\tensor*{A}{^I_a}} = \int_\sigma \mathrm{d}\tensor*{E}{^a_I} \wedge \mathrm{d}\tensor*{A}{^I_a} \qty(X_{f_1}, X_{f_2})=\int_\sigma \Big(\tensor*{\delta}{^a_1}\tensor*{\delta}{^1_I}\sin(r)^2\sin(\theta) + \tensor*{\delta}{^a_2}\tensor*{\delta}{^2_I}\sin(r)\sin(\theta)+ \\&\quad+ \tensor*{\delta}{^a_3}\tensor*{\delta}{^3_I}\sin(r)\Big)\mathrm{d}p \wedge \Big(\tensor*{\delta}{^1_a}\tensor*{\delta}{^I_1} + \tensor*{\delta}{^2_a}\tensor*{\delta}{^I_2}\sin(r) + \tensor*{\delta}{^3_a}\tensor*{\delta}{^I_3}\sin(r)\sin(\theta)\Big)\mathrm{d}c \,\qty(X_{f_1}, X_{f_2})\\
    &= \frac{3 V_0}{\sqrt{p}^3} \big\{f_1, f_2\big\}_{(p,c)} .
\end{align*}

In accordance with the following simplifying assumption, we will also simplify our seeting by using this symplectic structure as the reduced phase space of the discretised theory.\footnote{
We note that in the earlier literature also another submanifold of the phase space was investigated, namely in \cite{SKL06,APSV07}
\begin{align}
A^I_a(x)= c\;\omega^I_a(x) ,\hspace{30pt} E^a_I(x)= p\; \tilde{\omega}^a_I(x)
\end{align}
with $\omega$ the Maurer-Cartan form on $\mathfrak{su}(2)$.\\
Both are different submanifolds and interestingly when reducing of classical GR to these manifolds the evolution stays inside them. Therefore both description are classically equivalent. Whether both (or any of them) are also invariant submanifolds of discretised GR for the evolution produced by the full lattice Hamiltonian remains to be investigated.\\
For the purpose of this article we stay with the choice (\ref{classical_connection}) and (\ref{classical_triad}) as in this framework it is easier to compute the regularised Hamiltonian.}\\
To compute the evolution of a system in terms of two variables $(p,c)$ is sufficiently simple to be handleable. But in order to justify this, one must ask, what the relation between Poisson brackets in the full theory and at the reduced level is:
\begin{align*}
\{ F(h(e),E(e)), G(h(e),E(e)) \}_{(h,E)} |_{h,E \to c,p} \; \overset{{\bf ?}}{=}\; \{ F(h(e),E(e)) |_{h,E \to c,p}, G(h(e),E(e)) |_{h,E \to c,p} \}|_{(p,c)}
\end{align*}
where $h,E \to c,p$ refers to the symplectic reduction specified above. \\
A first, dissatisfactory observation is that both sides are in general {\it not} equal. This is the case only for special submanifolds $\, \overbar{\mathcal M}$ and sufficiently adapted functions $F,G$ (one of which typically needs to be invariant with respect to the symmetry of the system). An example for this is the Thiemann regularisation of the Hamiltonian on cubic lattices for isotropic, flat cosmology (see \cite{DKL20} for further details). However, we will in the following {\it assume} that symmetric reduction and Poisson brackets do commute in our situation, i.e. the above equation holds with an ``='' sign!\\
On the one hand this allows to simplify the computation of (\ref{Lorentzian}) drastically and on the other hand this assumption is anyway necessary if one wants to use the resulting expression as an effective Hamiltonian to generate evolution  with respect to some scalar field as we will do in the following section.\footnote{It is worth noting, that -- albeit not explicitly stated -- variants of this assumption are used in all effective models of LQC type once the evolution due to some reduced effective Hamiltonian is computed.} It is this conjecture which claims that the system at a later point of the flow induced by the constraints is still of the form of the computed discretisation of $k=+1$ cosmology.\\

With these assumptions we can simplify the evaluation of the formulas (\ref{Euclidian}) and (\ref{Lorentzian}) for isotropic, closed cosmology. However, the terms involved become quite lengthy and exceed the possibility to be printed as analytical results in a written paper. Therefore, we equate all $\epsilon_i=\epsilon$  and make a power series expansion for small $\epsilon$ to compute the leading orders of the corrections terms due to regularisation. The classical order is proportional to $\epsilon^0$ and to get some understanding of the behaviour of the evolution, we present the expansion up to order $\epsilon^7$. We receive for the Euclidian part
\begin{align*}
&C^\epsilon_E[1] |_{ h,E \to c,p} = \frac{\pi  (c^2-1) \sqrt{p}}{\kappa } \bigg(6 \pi-24 \epsilon +\Big(\frac{24}{\pi }-\frac{1}{8} \pi  (12 c^2+23)\Big) \epsilon ^2 +\Big(\frac{59 c^2}{9}+\frac{197}{18}\Big) \epsilon ^3+\\&+\frac{5456 \pi ^2 c^4+24 (1309 \pi ^2-10240) c^2+16641 \pi ^2-359040 }{34560 \pi }\,\epsilon ^4-\frac{32136 c^4+166828 c^2+68601 }{43200}\,\epsilon ^5\\&-\frac{544800 \pi ^2 c^6+16 (406933 \pi ^2-3107328) c^4+4 (3038605 \pi ^2-59329536) c^2+746287 \pi ^2-72571968 }{58060800 \pi }\,\epsilon ^6\\&+\frac{4786800 c^6+54461528 c^4+90626198 c^2-10179481}{101606400}  \,\epsilon ^7\bigg) +\order{\epsilon^8}
\end{align*}

and for the Lorentzian part a similar expansion is possible:
%
%
\begin{align*}
&(C^\epsilon[1]-C^\epsilon_E[1])|_{ h,E \to c,p} = \frac{(\frac{1}{\beta ^2}+1) c^2 \sqrt{p}}{\kappa }\Bigg(-6 \pi^2 +96 \pi  \epsilon+\bigg(\pi ^2 (6 c^2+\frac{53}{8})-672\bigg) \epsilon ^2\\&+\bigg(\frac{2688}{\pi }-\frac{17}{18} \pi  (104 c^2+109)\bigg) \epsilon ^3+\bigg(\frac{7}{9} (904 c^2+899)-\frac{\pi ^2 (84576 c^4+286264 c^2+83181)}{34560}-\frac{6720}{\pi ^2}\bigg) \epsilon ^4\\&+\bigg(\frac{10752}{\pi ^3}-\frac{98 (88 c^2+83)}{3 \pi }+\frac{\pi  (884328 c^4+2871482 c^2+746675)}{21600}\bigg) \epsilon ^5+\bigg(\frac{280 (236 c^2+211)}{9 \pi ^2}-\frac{10752}{\pi ^4}\\&-\frac{14545612 c^4+45330103 c^2+10444255}{48600}+\frac{\pi ^2 (10618880 c^6+76366776 c^4+90516776 c^2-3623677)}{19353600}\bigg) \epsilon ^6\\&+\bigg(-\frac{56 (1928 c^2+1633)}{9 \pi ^3}+\frac{20230744 c^4+60532486 c^2+12204505}{16200 \pi }\\&-\frac{\pi  (8566725504 c^6+59569235284 c^4+66493195964 c^2-4831037847)}{914457600}+\frac{6144}{\pi ^5}\bigg) \epsilon ^7  \Bigg)+\order{\epsilon^8}
\end{align*}

In context of effective LQC programme, this expression corresponds now to the approximated Hamiltonian of an isotropic, closed Universe, modified by the discreteness corrections, that emerge due to the Thiemann regularisation. In the next section, we will analyse these corrections in the regimes where the approximation to seventh order is justified, as well as compare it to the full regularised Hamiltonian.

\section{Numerical Analysis of the Approximation of the effective Hamiltonian constraint}
\label{s3}

In this section, we perform preliminary steps towards investigating the effective scalar constraint for $k=+1$ cosmological spacetimes of the previous section. For that purpose, we will proceed as in \cite{SKL06,APSV07} by coupling a massless, homogeneous scalar field to the geometry degrees of freedom. This scalar field will serve the role as a physical clock, i.e. since the flow of phase space parameters $(p,c)$ induced by the effective scalar constraint $C^\epsilon[N]$ is physically meaningless, we will need to compare it with the simultaneous flow of the scalar field $\Phi$ to deduce how physical quantities change with respect to each other. The total scalar constraint studied in this section reads:
\begin{align}
C_{\rm tot}[N]:= N\;C^\epsilon|_{h,E \to c,p} + N  \frac{\pi_\Phi^2}{2\sqrt{p^3}}
\end{align}
with some homogeneous lapse function $N$, and scalar field momentum $\pi_\Phi$ (canonically conjugated to $\Phi$, i.e. $\{\pi_\Phi, \Phi\}=1$). Throughout this section we work in natural units, i.e. $\ell_P=\hbar=G=c=1$.\\
Several remarks are in order:
\begin{itemize}
\item The lapse function $N$ is assumed to be homogeneous to respect the symmetries of the system. Since it only changes the unphysical flow of the scalar constraint and has no physical relevance, we will set $N\equiv1$ in the following.
\item  The matter part of the constraint incorporated here carries no further knowledge of the discretisation. This is not completely consistent as the matter Hamiltonian should be discretised as well. In principle, when inverse powers of the volume appear they should also be lifted via a version of Thiemann identities \cite{AQG1} to terms involving Poisson brackets, which upon regularisation have nontrivial contribution. However, to go along these lines would also require a consistent discretisation of the mater degrees of freedom, which has been omitted here as well. Therefore, this system serves merely as a toy model.
\item As was already stressed in the previous section, we make heavy use of the {\it assumption} that computing the flow on the reduced phase space agrees with the reduction of the flow on the full phase space. One should take note that the analytic result of the approximation to $C^\epsilon|_{h,E\to c,p}$  are also computed using said assumption. (The situation would get even more complicated if one would additionally incorporate the matter field phase space in the continuum and perform a discretisation and reduction of the total scalar constraint afterwards.)
\item Under the previous assumption the flow of $C^\epsilon |_{h,E \to c,p}$ would drive a phase space point parametrised by $(c(0), p(0), \Phi(0), \pi_\Phi(0))$ to a different point on the same reduced submanifold, i.e. $(c(t), p(t), \Phi(t),$ $\pi_\Phi(t))$ to allow for relational observations. Concerning the analytical, approximated constraints we must moreover isolate the physical viable regions.
This means, we must carefully investigate a priori at which points in the reduced phase space the approximation to $7 ^{\rm th}$ order is valid, and whether the flow leaves at some point this regime of validity. Points in phase space where the approximation breaks down and any effects found thereon carry {\it no} physical relevance.
\end{itemize}

\subsection{Preliminary Analysis}

We will start our analysis by fixing the free parameters of the model, i.e. $\epsilon$ and $(c(0), p(0), \Phi(0), \pi_\Phi(0))$, such that they allow for a sufficiently classical regime. The choice of the regulator $\epsilon$ refers to the lattice spacing with respect to coordinate distance, which we choose earlier to be the same in each direction $r, \theta, \varphi$ \footnote{For the purpose of this article, we will not comment on the fact that $\epsilon\in\re$ corresponds to what is called $\mu_o$ scheme in the LQC literature. It is apparent that due to the reference to a fiducial coordinate system, the physical predictions get affected by coordinate effects.  Several preliminary proposals exist in the literature of how this can be remedied, e.g. making $\epsilon$ change under scaling as well. The most prominent of them is the $\bar\mu$ -scheme introduced in \cite{APS06c} and frequently used in the literature. However, until to today there is no satisfactory derivation of the $\bar\mu$ scheme from the full theory and therefore we refrain from using it.}. We choose our coordinate system such that $r\in[0,\pi/2)$, $\theta\in [0,\pi)$ and $\varphi\in [0,2\pi)$. The lattice spacing $\epsilon$ as measured in said coordinate distance must therefore be sufficiently smaller than $\pi/2$ in order to allow for an acceptably dense graph. Following (\ref{vertices_epsilon}) a possible choice could be $\epsilon=\pi/100$, corresponding to a lattice with $10^6$ many vertices. We point out that this is in contrast to the standard $\mu_0$ scheme of LQC literature, where the regularisation parameter is commonly chosen to be $\mu_0=3\sqrt{3}$ (in Planck units) by relating it to the minimal eigenvalue of the area operator in LQG \cite{APS06a}. However, such a value exceeds the coordinate spacing of our manifold and we want to keep its natural value -- indicating that we should refrain from choosing the $\mu_0$-regularisation. There is also a strong reason, why such a choice is disfavoured for the current scalar constraint:\\

We are dealing with $k=+1$ cosmology and want to investigate solutions that feature a classical behaviour. Consequently, we search for those trajectories in phase space that have a {\it recollapse point}  in the classical regime. A recollapse point in the phase space is defined by lying in the hyperplane $c=0$.\footnote{In standard GR, this automatically equated to $\dot{p}=0$. Therefore implying a change from an expanding to a collapsing universe.} In order for the system to behave classical, the energy density $\rho:= \pi_\Phi^2/(2\pi^2p^{3})$ must be a positive value. To see, what kind of restrictions this poses, let us look at the expansion of $C^\epsilon_{\rm tot}$ up to first order in $\epsilon$:
\begin{align}\label{recollapse_first_order}
C^\epsilon_{\rm tot} (c=0,p,\Phi,\pi_\Phi)=\frac{\pi_\Phi^2}{2p^{3/2}}-\frac{6\sqrt{p}\pi^2}{\kappa}+\frac{24 \sqrt{p}\pi \epsilon}{\kappa}+\mathcal{O}(\epsilon^2)
\end{align}
Upon imposing vanishing of the constraint, and demanding positivity of $\rho_c$ it follows that: $(6\pi - 24\epsilon) > 0$. In other words, a classical recollapse can only occur if $\epsilon < \pi/4 <\mu_o$.\\
Of course, this is only for the first order expansion of the constraint -- if one evaluates $C^\epsilon$ numerically at $c=\pi_\Phi=0$ one finds the requirement: $\epsilon< \epsilon_{\rm max} = 1.19873$ (which is slightly bigger than $\pi/4\approx 0.7853$).
While this requirement eliminates the possibility of choosing the $\mu_o$ scheme, it does of course not fix a minimal value for $\epsilon$. Thus, we are free to take for the upcoming investigations $\epsilon=\pi/100$.\\

However, these modifications proportional to $\epsilon$ go with the same power of $p$ as the classical curvature term. This implies that even around the recollapse, where the universe is mostly classical, a possible deviation from standard GR directly proportional to $\epsilon$ could be measured, if one assumes the regularisation $C^\epsilon_{\rm tot}$.\\
For the range of allowed values for the regularisation parameter $\epsilon\in [0,\epsilon_{\rm max}]$ one can check that the derivations from the reduced scalar constraint of classical GR are always positive. Therefore, in order to satisfy the constraint for fixed $\pi_\Phi$, we will always need a $p_{\rm disc}$ in the discrete model that is larger than the square of the scale factor for classical general relativity, i.e. $p_{\rm clas} < p_{\rm disc}$ at the recollapse.\\
Such a behaviour was never encountered in LQC before, due to the fact that all modifications proportional to $\epsilon$ appeared as functions of the form $f(c\epsilon)$, i.e. in the ``limit of late time cosmology'' $c\to 0$ any modification proportional to $\epsilon$ would be damped as well. This model therefore presents a novel feature, that quite generally can happen for any discretisation.\\
It is interesting to note that the modification due to $\epsilon$ for the scalar constraint at the recollapse, could be absorbed in a redefinition of $\kappa$. E.g. for (\ref{recollapse_first_order}) (which only includes linear corrections in $\epsilon$) the dynamics in the classical regime can be equivalently described by the scalar constraint of standard GR with the effective Newton constant: $G_{\rm eff}:=G (1-4\epsilon/\pi)$. Comparing measurements of Newtons constant from cosmology and other methods could therefore provide a further, upper bound for $\epsilon$.\\

\begin{figure}[H]
	\begin{center}
	\includegraphics[scale=0.46]{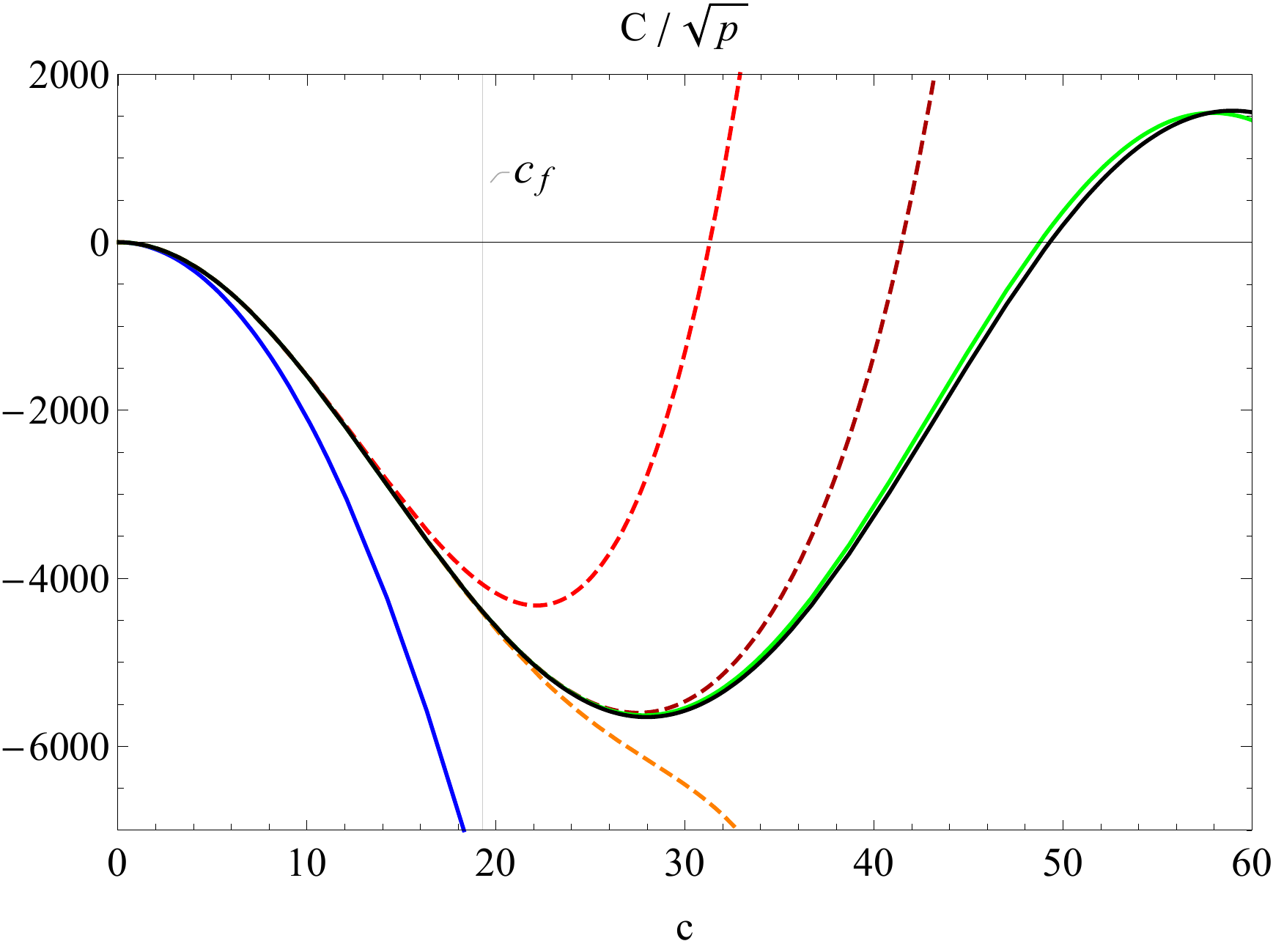}
		\includegraphics[scale=0.46]{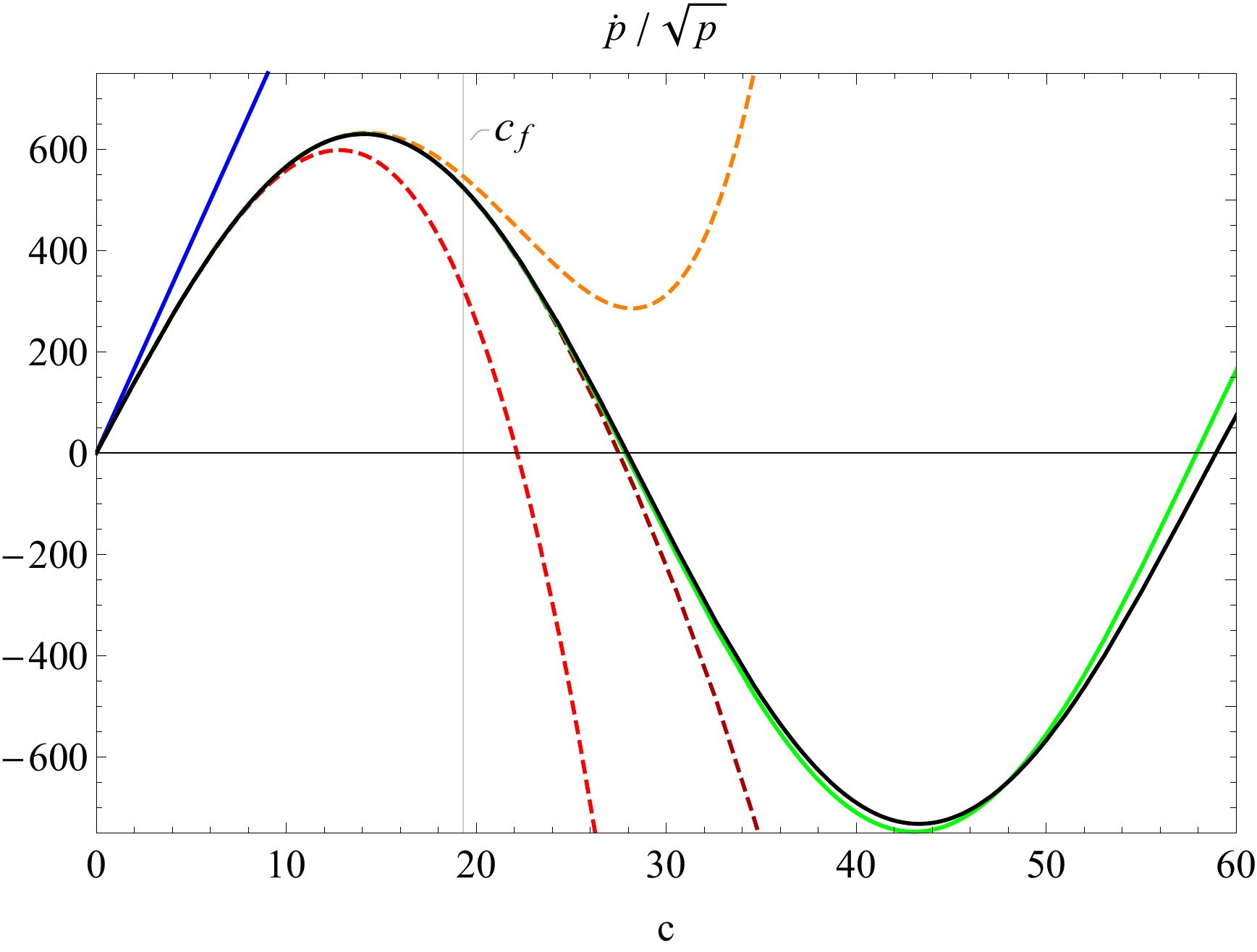}
	\end{center}
\caption{\it{Plot of $c$ vs $C^\epsilon/\sqrt{p}$ (left) and vs $\dot{p}/\sqrt{p}$ (right) as obtained from Hamiltons equation using different scalar constraints for $\beta=0.2375$ and $\epsilon=\pi/100$. In blue, the evolution was obtained from $C$ of standard GR; in red, dashed the $3^{\rm rd}$ order approximation in $\epsilon$ of $C^\epsilon_{\rm tot}$, in orange, dashed  the $5^{\rm th}$ order approximation in $\epsilon$ of $C^\epsilon_{\rm tot}$ and in dark red, dashed the $7^{\rm th}$ order approximation in  $\epsilon$ of $C^\epsilon_{\rm tot}$ are presented. Finally, we show in black the numerical evaluation of $C^\epsilon$ for the given values of $c$ and in green the proposed effective model from (\ref{effective_model}). As seen from analytical arguments, $C^\epsilon_{\rm tot}$ forces $c$ to strictly increase monotonic in backward time evolution. The crossing through zero of $\dot{p}/\sqrt{p}$ for finite values of $c$ implies a bounce of the model. Evaluating backwards from the classical recollapse point, the model will feature exactly one asymmetric bounce. Due to the symmetric behaviour of the constraint in $c$ the values are point-symmetrically mirrored to negative values of $c$ describing the flow in forward time evolution.} \label{fig1}}
\end{figure}

The choice of $\Phi$ and $\pi_\Phi$ are independent of the choice for $\epsilon$ at the recollapse. Via several numerical simulations, we found out that the qualitative behaviour of the phase space trajectories does not change for different values of $\pi_\Phi$. The latter one is a constant of the dynamics, since $\Phi$ is a cyclic variable, i.e. it does not appear in $C^\epsilon_{\rm tot}$. Therefore, the absolute value of $\Phi$ is irrelevant as well and merely corresponds to a time shift. Once a value of $\pi_\Phi$ is specified, by imposing the constraint at the recollapse, one can determine $p$ and start numerical simulations.\\
However, before doing so, we must determine a priori until what point the flow of $C^\epsilon_{\rm tot}$ agrees at least qualitatively with the expansions to which we have analytical access. For that purpose, we investigate Hamilton's equation of motion analytically. First we determine $\dot{c}$, which, after imposing vanishing of the constraint, reads:
\begin{align}\label{cdot}
\dot{c} =\{ C^\epsilon_{\rm tot} , c\} = \frac{\kappa \beta}{6 \pi^2} \frac{\partial C^\epsilon_{\rm tot}}{\partial p} \overset{C^\epsilon_{\rm tot}=0}{=} \frac{\kappa \beta}{6 \pi^2}\bigg[(-3/2)\frac{\pi_\Phi^2}{2 p^{5/2}}+ (-1/2)\frac{\pi_\Phi^2}{2 p^{5/2}}\bigg]= - \frac{\kappa \beta}{6 \pi^2}\frac{\pi_\Phi^2}{p^{5/2}}
\end{align}
This implies that $c$ decreases strongly monotonic, since we have chosen positive orientation of the triad (in agreement with positive volume $p>0$). Especially, the change in $c$ grows the closer the flow drives towards a singularity.\\
Further, we plot $\dot{p}/\sqrt{p}$ in \autoref{fig1}, where the time derivative is obtained from Hamilton's equation, i.e.
\begin{align}
\dot{p}=\{ C^\epsilon_{\rm tot}, p\}= \sqrt{p} \;\; {\rm Fun}(c,\epsilon,\beta),\hspace{30pt}
\dot{p}_k = \{C^k, p\} = \sqrt{p}\;\; {\rm Pol}_k(c,\epsilon,\beta)
\end{align}
where $C^k$ denotes the expansion of $C^\epsilon_{\rm tot}$ to order $\epsilon^k$ and ${\rm Fun}$ denotes a function to which we have only numerical access, while ${\rm Pol}_k$ are known in closed form. It transpires from \autoref{fig1} that there exist points in phase space - namely at high absolute values of $c$ - where a strong deviation of the seventh order from the numerically known exact data occurs. This indicates a regime where one should no longer trust the expansions. Since \autoref{fig1} is presenting $\dot{p}$ one sees that this deviation appears around the bounce (i.e. the point where $\dot{p}(c)=0$).
Due to the symmetric form of the constraint (it depends only on natural powers of $c^2$) the same effects appear also in forward evolution.\\

In addition to the $N^{\rm th}$ order approximations, we therefore propose an {\it effective} model for $\epsilon=\pi/100$: We postulate the following effective scalar constraint:
\begin{align}\label{effective_model}
C_{\rm eff}^\epsilon=\frac{6\pi^2\sqrt{p}}{\kappa}\left(\frac{\sin(\alpha \,\epsilon \,c)^2}{\alpha^2 \epsilon^2}\alpha_2-[1-\frac{24\pi\epsilon}{\kappa}+\frac{9\pi \epsilon^2}{\kappa}]-\frac{1+\beta^2}{\beta^2}\frac{\sin(2\, \alpha\, \epsilon\, c)^2}{4\, \alpha^2 \epsilon^2}\alpha_3\right)
\end{align}
with $\alpha=0.867-0.116\epsilon,\; \alpha_2=1-1.231\epsilon, \; \alpha_3= 1-4.97\epsilon+6.846\epsilon^2$.\\
These values have been chosen in such a way that the Euclidian and Lorentzian part provide a qualitative fit to the numerical exact data. In \autoref{fig1} we present the resulting $\dot{p}/\sqrt{p}$ equation in green - its agreement with the black curve (the numerical data) is very good even for very early/late times. Let us note that $C^\epsilon$ and $C_{\rm eff}^\epsilon$ both feature a transition through zero around $c\approx 50$, whence satisfaction of the constraint (including matter) implies that for $c\to 50$ we find $p\to \infty$. That is, a big rip occurs before further values of $c$ are reached and thus while the quality of $C_{\rm eff}^\epsilon$ decays in this part of the phase space, it is irrelevant from a physical perspective.\\
Finally, we point out that the effective model is of course not a perfect fit to the numerical data, especially around the recollapse phase, the $7^{\rm th}$ order approximation yields better results. Therefore, we denote the point where $\dot{p}_7$ to $\dot{p}_{\rm eff}$ deviate in the same amount from the numerical data by $c_f\approx 19.3$: in the regime $[0,c_f]$ the $7^{\rm th}$ order is the best approximation to the numerical data, while in $[c_f,50]$ the effective model is superior.\\

We can draw further analysis from \autoref{fig1}. Namely, due to (\ref{cdot}) it is apparent that $c$ increases strongly monotonic in backward time evolution into the far past. Then, the $\dot{p}(c)$ diagram of \autoref{fig1} indicates that at the transition through zero a bounce can occur. However, if this happens no second recollapse point can be present as the diagram of $C^\epsilon(c)$ features for $c\approx 50$ a transition to zero, which -- upon taking matter into account -- indicates that $p\to \infty$ at said transition, therefore indicating that the universe expands to infinite volume in backward time evolution. In other words, the bounce predicted by the model must happen in an {\it asymmetric} fashion!

\subsection{Numerical Simulations}

\begin{figure}[H]
	\begin{center}
		\includegraphics[scale=0.75]{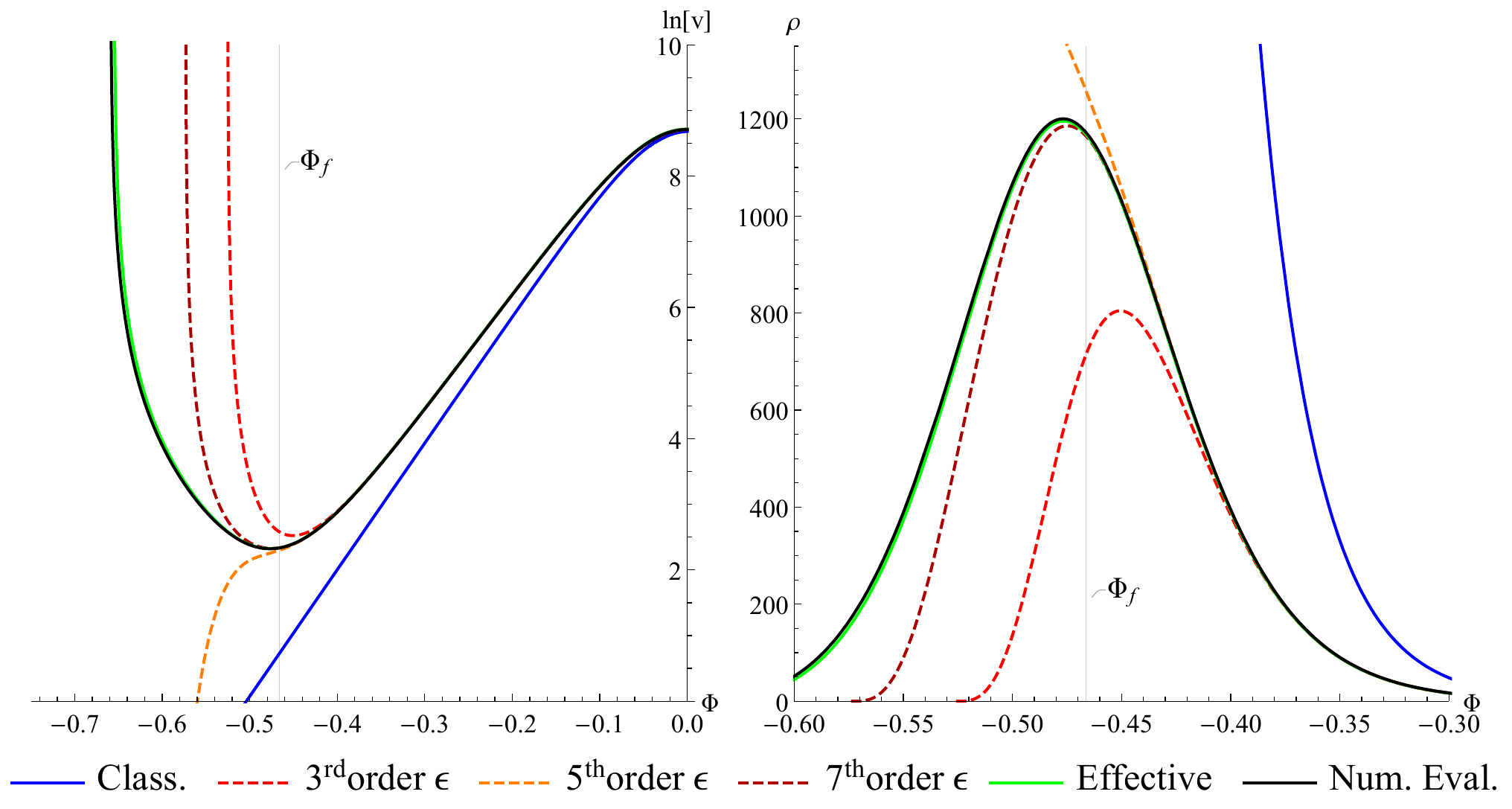}
	\end{center}
\caption{\it{Case (A) : $\pi_\Phi=500$. Flow of volume $v=\pi^2 p^{3/2}$ and energy density $\rho= \pi_\Phi^2/(2\pi^2 p^{3/2})$ as driven by various constraints. In blue, from C of standard GR; in red, dashed the $3^{\rm rd}$ order approximation in $\epsilon$ of $C^\epsilon_{\rm tot}$, in orange, dashed  the $5^{\rm th}$ order approximation in $\epsilon$ of $C^\epsilon_{\rm tot}$ and in dark red, dashed the $7^{\rm th}$ order approximation in  $\epsilon$ of $C^\epsilon_{\rm tot}$ are presented. 
The numerical evaluation of $C^\epsilon$ is shown in black and in green the effective model (\ref{effective_model}) capturing the qualitative features.
The initial data were picked at the recollapse at $\Phi(0)$. One finds an asymmetrical bounce transitioning into an super-exponentially contracting universe. Due to the symmetry of the constraints, the flow is mirrored in positive $\Phi$ direction. \label{fig2}}}
\end{figure}

\begin{figure}[H]
	\begin{center}
	\includegraphics[scale=0.75]{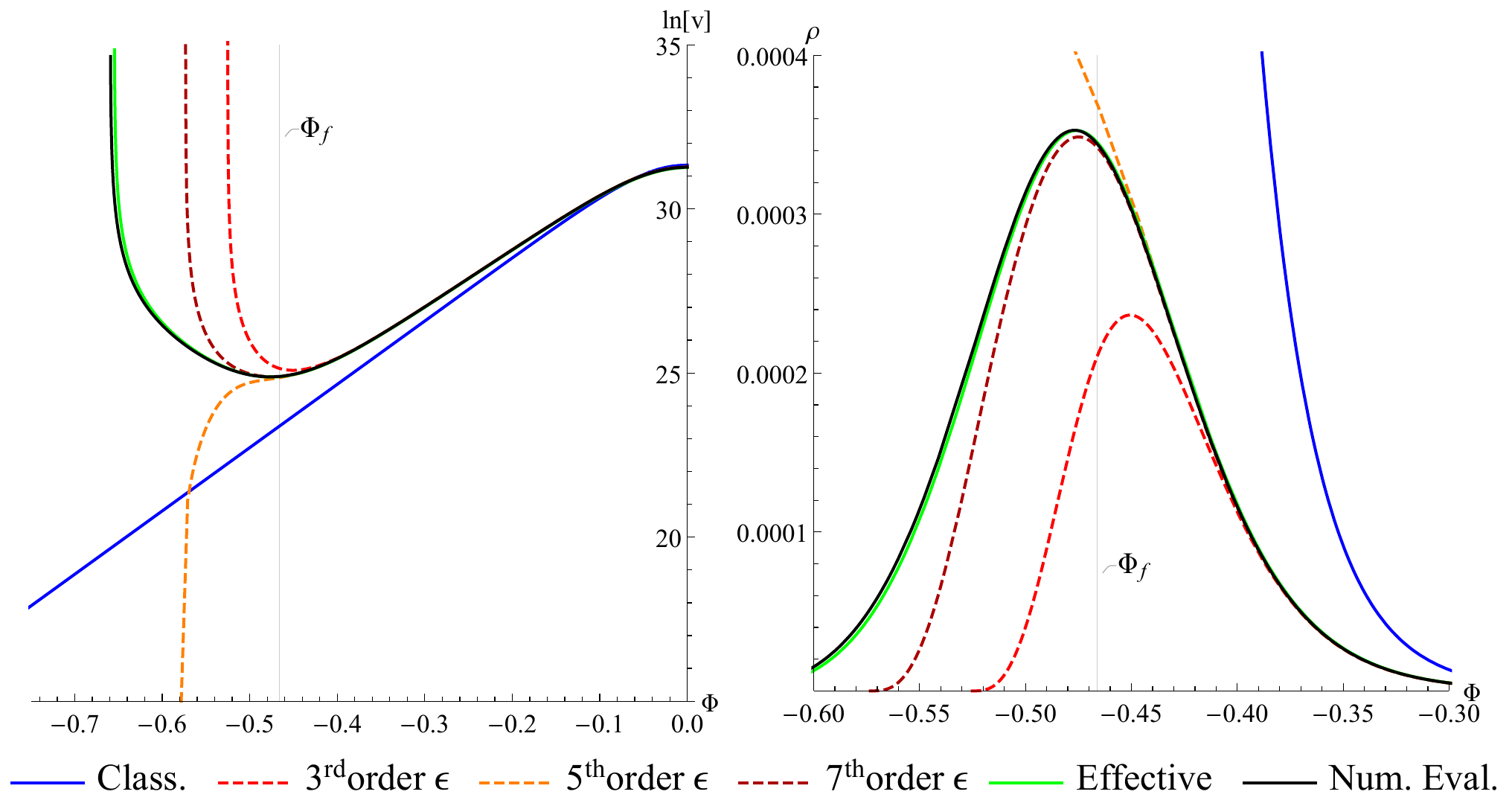}
	\end{center}
\caption{\it{Case (B) : $\pi_\Phi=1.77\times 10^9$. Flow of volume $v=\pi^2 p^{3/2}$ and energy density $\rho= \pi_\Phi^2/(2\pi^2 p^{3/2})$ with respect to scalar field time $\Phi$ as driven by various constraints. The notations are the same as in \autoref{fig2}. \label{fig3}}}
\end{figure}

Now, we will study the flow of the constraint by numerical methods for observables volume $v=\pi^2 p^{3/2}$ and energy density $\rho= \pi_\Phi^2 /(2\pi^2p^3)$ . We stress again that we assume validity of replacing the symplectic structure of the discretised phase space with the reduced  Poisson bracket $\{p,c\}=\beta\kappa/(6 \pi^2)$. For the Immirzi parameter we take $\beta=0.2375$ as is custom in LQC literature. As initial data, we pick the recollapse at $c(0)=0$ with the arbitrary choice $\Phi(0)=0$ and determine $p(0)$ via imposing the respective constraint. It remains therefore to choose the constant of motion $\pi_\Phi$. We will present two cases: (A) for $\pi_{\Phi,A}=500$ and (B) $\pi_{\Phi,B}=1.77\times 10^9$, and it will transpire that the qualitative behaviour of the phase space trajectory is unaffected by this choice (The later value corresponds to a choice where $\rho=10^{-9}$ at the recollapse). We plot in \autoref{fig2} and \autoref{fig3} the flow of the $v$ and $\rho$, deparametrised with respect to scalar field time $\Phi$, for classical GR constraint and numerical $C^\epsilon_{\rm tot}$, as well as the analytical approximations of $C^\epsilon_{\rm tot}$, namely for $3^{\rm rd}$, $5^{\rm th}$ and $7^{\rm th}$ order expansion in $\epsilon$. Moreover, we show the evolution of the effective model (\ref{effective_model}), which stays qualitatively close to the evolution predicted by $C^\epsilon_{\rm tot}$. However, around the recollapse the $7^{\rm th}$ order is closer to the numerical approximation. The borders of this regime are marked by $\Phi_f:=\Phi|_{c=c_f}$ (see previous section).
One can deduce that close to the initial/final (due to time reflection symmetry of the model) singularities a deviation from standard general relativity takes place. To be precise in the early universe we would expect that the expansion was not as strong with respect to $\Phi$ as predicted by classical GR. Instead, an asymmetric bounce presents a transition from/to another universe, which underwent a super-exponential contraction/expansion. These universes however do not present another recollapse point, instead reach infinite volume in finite $\Phi$-time. The effect is driven by the presence of the Lorentzian term as can be seen from the effective model which mimics the qualitative features and is presents therefore a striking similarity to other models which took regularisations of the Lorentzian part into account \cite{ADLP19}. \\

We shall also comment on possible violations of the approximated constraint: Independent of the chosen value for $\pi_\Phi$ the violation of the constraint will always be satisfied in the regime between recollapse and $\Phi_f$. However, before $\Phi_f$, before the ``bounce'' occurs the system features an exponential fast contraction -- $p$ decreases by 4 orders of magnitude in 0.6 $\Phi$-time. This causes a drastic violation of the constraint in absolute values $C^\epsilon_{\rm tot} \neq 0$, however compared to the absolute value of the volume of the model this violation is still negligible. A similar situation happens in the LQC cases of cosmological constants that feature a similar exponential contraction/expansion and where only $C_\Lambda /v$ remains small.\\

Finally, it is also interesting to note that whether a bounce happens or not is not an intrinsic features that only occurs if the constraint is bounded with respect to $c$ (e.g., as is the case if $c$ appears only inside of trigonometric functions). Instead, also truncations of finite powers of trigonometric functions can cause a bounce (we see that a bounce is predicted for $3 ^{\rm rd}$ and $7^{\rm th}$ order). However, a bounce is also not a unique criterion that always appears as soon as $\epsilon$-corrections a present the $5 ^{\rm th}$ order is driven into a singularity -- this happens necessarily as also \autoref{fig1} showed that the corresponding $\dot{p}$-equation never crosses zero, i.e. there exists no turning point.

\subsection{Comparison to earlier closed LQC models}
In previous literature, a LQC-like quantisation on a reduced phase space for $k=+1$ models was already proposed. There, an effective constraint was derived from the reduced quantum theory which serves as regularisation of the $k=+1$ -version of the scalar constraint \cite{SKL06,APSV07}. However, a crucial ingredient in this construction was the following identity which is {\it only true for certain coordinates in $k=+1$ spacetime} \cite{SKL06}:
\begin{align}\label{simplf}
2K^I_{[a} K^J_{b]} =\frac{1}{\beta^2}\epsilon^{IJ}_{\;\; K} F^K_{ab}+\frac{1}{2}\;^0\omega^I_{[a} \;^0\omega^J_{b]}
\end{align}
where $^0\omega$ is the Maurer-Cartan form on $\mathfrak{su}(2)$.\\
Using this simplification which only holds true in this closed isotropic models, a much simpler regularisation of the scalar constraint can be obtained (i.e. it avoids a lengthy regularisation of the Euclidean part). The philosophy behind this procedure can be rephrased as {\it first reducing and then discretising} and as one can see these two procedures do not commute (this was already observed in isotropic flat cosmology).\\
In \cite{APSV07} moreover a further simplification was assumed, namely that the holonomy over {\it any} plaquette in the $k=+1$ spacetime has the same functional form in terms of the parameteres $c,p$. This is -- of course -- not the case in general and also not for the coordinates of $k=+1$ spacetimes used neither here nor in \cite{APSV07}. Therefore, even is one would employ (\ref{simplf}) before discretisation, afterwards performing the smearing over the whole spatial manifold would lead to a different effective Hamiltonian.\\
Nonetheless, we want to point out that this is no caveat of the methods of \cite{SKL06} and \cite{APSV07}: the authors implement moreover the so-called $\bar{\mu}$-scheme \cite{APS06c} a posteriori by simply replacing $\epsilon \to \bar{\mu}\propto 1/\sqrt{p}$. Since there is up to today no quantisable regularisation of the Hamiltonian in  full general relativity known that features a $\bar{\mu}$-scheme (see \cite{DLP19} for a discussion of the case of flat cosmology and \cite{Bod15,HL19_Improv} for first steps towards its implementation)  any relation to the Thiemann-regularisation is anyway far-fetched. However, the $\bar{\mu}$-scheme resolves an issue regarding the remnant of residual diffeomorphisms. In this sense, if one does not require that the effective Hamiltonian stems from a valid regularisation of the full theory but focuses on solving the rescaling problem, the proposal of \cite{SKL06} and \cite{APSV07} presents a very successful candidate.

\section{Conclusion}
\label{s4}
In this paper we investigated the dynamics of discretised general relativity modelling a closed, spatially isotropic universe. We proposed as fundamental spatial manifold a family hyperspherical lattices embedded in $S^3$, on which the metric degrees of freedom are encoded in terms of the holonomies and fluxes of the Ashtekar-Barbero framework on the edges of the lattice. Such a discretisation allows for an approximation of isotropic spatial data (which cannot be exact as the discrete data cannot remain invariant under arbitrary rotations and translations of the $S^3$). The scalar (or Hamiltonian) constraint $C$ of general relativity is discretised, i.e. replaced by a function $C^\epsilon$ expressed in terms of quantities on the lattice which only agrees with $C$ in the limit of infinitely dense lattices. The functional form of $C^\epsilon$ chosen in this paper is the Thiemann regularisation which is custom in LQG, justifying the expectation that our computation could capture certain aspects of a theory of quantum gravity.\\
Due to the complicated structure involved, we approximate the analytical form of $C^\epsilon$ to finite order in terms of the lattice spacing, i.e. $7^{\rm th}$ order. With this, we improve earlier work in the literature by not requiring any symmetries of the system prior to discretisation, although we ,too, work under the assumption that the reduced dynamics agrees with the dynamics of the graph (i.e. under premise that the flow of $C^\epsilon$ does not kick an initially, approximate isotropic configuration out of its subspace). The flow of these approximated constraints can then be studied on the reduced level and be compared to the numerical evaluation of the full scalar constraint on the interesting regions of the phase space.\\
We have put these approximations to the test by coupling them to a massless free scalar field and studying the flow of this cosmological toy model with analytic and numerical tools: all the models feature a point where $c=0$, i.e. a classical recollapse point. Indeed, discretisation (or potentially quantum) effects are suppressed at this point to feature largely the known dynamics. However, a small imprint of the discretisation remains that can be absorbed into a rescaling of Newtons coupling constant around the recollapse point. The deviation from the classical constant is directly proportional to the lattice spacing and could therefore in principle be used to find an upper bound for the lattice spacing. Further, since starting from the recollapse point the classical universe is driven towards initial and final singularities. In contrast, $C^\epsilon$  - which can be evaluated numerically - features instead a corresponding big bounce in each direction, bridging to exponentially fast contracting/expanding universes. We compared also with the evolution of the approximations, for which the $7^{\rm th}$ order presents a qualitative close behaviour to the exact numerical data almost until the bounce. In principle, this makes the approximation very interesting for in-depth analysis of physical effects in the universe close to the bounce. Finally, we also presented an {\it effective model} which mimics also the late time behaviour of $C^\epsilon$, while being of a much simpler analytic form. Interestingly, the effective model (\ref{effective_model}) is in its form very similar to the classical model, by majorly replacing the connection with a polymerisation: $ c \mapsto \sin (\alpha c)$. However, the exact form of the frequency $\alpha$ could not have been guessed a priori.\footnote{ The fact, that the huge summation of the lattice boils down to such a simple expression also strengthens strategies to compute effective models such as the one presented for black holes in \cite{ADL19}.}\\

In future work, it will be of interest to increase the resolution around the singularity, e.g. by computing higher order approximations of the effective scalar constraint. Also further investigations of the effective model are interesting and to determine whether such effective models can also be found for more complicated systems. It is also an open question, to what degree different graphs as underlying discretisation of the spatial manifold will have impact on the effective dynamics of cosmological solutions. Finally, any serious investigations of the flow requires further work to justify the assumptions of reducing the dynamical evolution to the reduced sector. Promising work, asking for the conditions under which this is possible, is ongoing \cite{DKL20}.

\section*{Acknowledgements}
The authors thank Andrea Dapor, Kristina Giesel and Parampreet Singh for discussions.
This work is partially supported by NSF grant PHY-1454832 and by the project BA4966/1-2 of the German Research Foundation (DFG). KL also acknowledges support by the DFG under Germany's Excellence Strategy - EXC 2121 ``Quantum Universe'' - 390833306.


\begin{thebibliography}{99}
{\setlength{\parskip}{0.0em}		\setlength{\itemsep}{0.6pt}

\bibitem{Fr22}
	A. Friedmann.
	``\"Uber die Kr\"ummung des Raumes".
	{\it Zeitschrift f\"ur Physik A} {\bf 10}
	(1922)
	
\bibitem{Fr23}
	A. Friedmann.
	``Die Welt als Raum und Zeit (The World as Space and Time)".
	{\it Ostwalds Klassiker der exakten Wissenschaften}
	ISBN 3-8171-3287-5
	
\bibitem{Fr24}
	A. Friedmann.
	``\"Uber die M\"oglichkeit einer Welt mit konstanter negativer Kr\"ummung des Raumes''.
	{\it Zeitschrift f\"ur Physik A} {\bf 21}
	(1924)

\bibitem{Le27}
	G. Lema\^{\i}tre.
	``Un univers homog\`ene de masse constante et de rayon croissant rendant compte de la vitesse radiale des nebuleuses extra-galactiques''.
	{\it Annales de la Societe Scientifique de Bruxelles} {\bf A47}
	(1927)

\bibitem{Le31}
	G. Lema\^{\i}tre.
	``Expansion of the universe, A homogeneous universe of constant mass and increasing radius accounting for the radial velocity of extra-galactic nebulae''.
	{\it Monthly notices of the royal Astr. Society} {\bf 91}
	(1931)

\bibitem{Le33}
	G. Lema\^{\i}tre.
	``L'Univers en expansion''.
	{\it Annales de la Societe Scientifque} {\bf A53}
	(1933)

\bibitem{Rob35}
	H.P. Robertson.
	``Kinematics and world structure".
	{\it Astrophysical Journal} {\bf 82}
	(1935)

\bibitem{Rob36a}
	H.P. Robertson.
	``Kinematics and world structure II".
	{\it Astrophysical Journal} {\bf 83}
	(1936)

\bibitem{Rob36b}
	H.P. Robertson.
	``Kinematics and world structure III".
	{\it Astrophysical Journal} {\bf 83}
	(1936)

\bibitem{Wal37}
	A.G. Walker.
	``On Milne's Theory of world-structure''.
	{\it Proceedings of the London Mathematical Society} {\bf 42}
	(1937)
	
\bibitem{Muk05}
	V. Mukhanov.
	``Physical Foundations of Cosmology''.
	{\it Cambridge University Press}
	(2005)

\bibitem{dS34}
	W. de Sitter.
	``On distance, magnitude and related quantities in an expanding universe".
	{\it Bulletin of the Astronomical Ins. of the Netherlands} {\bf 7}
	(1934)


\bibitem{GP00}
	R. Gambini, J. Pullin.
	``Loops, knots, gauge theories and quantum gravity".
	{\it Cambridge University Press}
	(2000)
	
\bibitem{Rov04}
	C. Rovelli.
	``Quantum gravity".
	{\it Cambridge University Press}
	(2004)
	
\bibitem{AL04}
	A. Ashtekar, J. Lewandowski.
	``Background independent quantum gravity: A Status report".
	{\it Class. Quant. Grav.} {\bf 21}, R53
	(2004)                    
	
\bibitem{Thi09}
	T. Thiemann.
	``Modern canonical quantum general relativity".
	{\it Cambridge University Press}
	(2008)


\bibitem{Ash86}
	A. Ashtekar.
	``New variables for classical and Quantum Gravity".
	{\it Phys. Rev. Lett.} {\bf 57}, 2244-2247
	(1986)
	
\bibitem{Ash87}
	A. Ashtekar.
	``New Hamiltonian formulation of General Relativity".
	{\it Phys. Rev. D} {\bf 36}, 1587-1602
	(1987)
	

\bibitem{Bar94}
	J.F. Barbero.
	``A real polynomial formulation of General Relativity in terms of connection".
	{\it Phys. Rev. D} {\bf 49}, 6935-6938
	(1994)
	
\bibitem{Bar95}
	J.F. Barbero.
	``Real Ashtekar Variables for Lorentzian Signature Space-times".
	{\it Phys. Rev. D} {\bf 51}, 5507-5510
	(1995)
	
\bibitem{Boj99a}
	M. Bojowald.
	``Loop Quantum Cosmology I: Kinematics''.
	{\it Class. Quant. Grav.} {\bf 17}, 1489-1508
	(2000)
	
\bibitem{Boj99b}
	M. Bojowald.
	``Loop Quantum Cosmology II: Volume Operators''
	{\it Class. Quant. Grav. } {\bf 17}, 1509-1526
	(2000)

\bibitem{ABL03}
	A. Ashtekar, M. Bojowald, J. Lewandowski.
	``Mathematical structure of loop quantum cosmology''.
	{\it Adv. Theor. Math. Phys.} {\bf 7}, 233-268
	(2003)

\bibitem{Boj05}
	M. Bojowald.
	``Loop Quantum Cosmology''.
	{\it Living Rev. Relativity} {\bf 8}, 11
	(2005)
	
\bibitem{Ash08}
	A. Ashtekar.
	``Loop Quantum Cosmology: An Overview''.
	{\it Gen. Rel. Grav.} {\bf 41}, 707-741
	(2009)
		


\bibitem{Engle07}
	J. Engle.
	``Relating loop quantum cosmology to loop quantum gravity: Symmetric sectors and embeddings''.
	{\it Class. Quant. Grav.} {\bf 24}, 5777-5802
	(2007)
	
\bibitem{Fle15}
	C. Fleischhack.
	``Kinematical Foundations of Loop Quantum Cosmology''.
	[arXiv:1505.04400]
	(2015)
	
\bibitem{BEHM16}
	C. Beetle, J. Engle, M. Hogan, P. Mendonca.
	``Diffeomorphism invariant cosmological symmetry in full quantum gravity''.
	{\it Int. J. Mod. Physics D} {\bf 25}, 1642012
	(2016)
	

\bibitem{Thi98a}
	T. Thiemann.
	``Quantum Spin Dynamics (QSD) I''.
	{\it Class. Quant. Grav.} {\bf 15}, 839-873
	(1998)
	
\bibitem{Thi98b}
	T. Thiemann.
	``Quantum Spin Dynamics (QSD) II''.
	{\it Class. Quant. Grav.} {\bf 15}, 875-905
	(1998)

\bibitem{AQG1}
	K. Giesel, T. Thiemann.
	``Algebraic Quantum Gravity (AQG) I. Conceptual Setup''.
	{\it Class. Quant. Grav.} {\bf 24}, 2465-2498
	(2007)
	
\bibitem{AQG2}
	K. Giesel, T. Thiemann.
	``Algebraic Quantum Gravity (AQG) II. Semiclassical Analysis''.
	{\it Class. Quant. Grav.} {\bf 24}, 2499-2564
	(2007)
	
	

\bibitem{AC13}
	E. Alesci, F. Cianfrani.
	``Quantum-reduced loop gravity: cosmology''.
	{\it Phys. Rev. D} {\bf 87}, 083521
	(2013)

\bibitem{AC14a}
	E. Alesci, F. Cianfrani.
	``Quantum Reduced Loop Gravity: Semiclassical limit''.
	[arXiv:1402.3155]
	(2014)

\bibitem{AC14b}
	E. Alesci, F. Cianfrani.
	``Loop Quantum Cosmology from Loop Quantum Gravity''.
	[arXiv:1410.4788]
	(2014)	
	
\bibitem{DL17a}
	A. Dapor, K. Liegener.
	``Cosmological Effective Hamiltonian from full Loop Quantum Gravity''.
	{\it Phys. Lett. B} {\bf 785}, 506-510
	(2018)
	
\bibitem{DL17b}
	A. Dapor, K. Liegener.
	``Cosmological Coherent State Expectation Values in LQG I. Isotropic Kinematics''.
	{\it Class. Quant. Grav. }{\bf 35}, 135011
	(2018)

\bibitem{ADLP18}
	M. Assanioussi, A. Dapor, K. Liegener, T. Pawlowski.
	``Emergent de Sitter epoch of the quantum Cosmos''.
	{\it Phys. Rev. Lett.} {\bf 121}
	(2018)

\bibitem{ADLP19}
	M. Assanioussi, A. Dapor, K. Liegener, T. Pawlowski.
	``Emergent de Sitter epoch of the quantum Cosmos: a detailed analysis''.
	{\it Phys. Rev. D} {\bf 100}, 084003 
	(2019)
	
\bibitem{GM19a}
	A. Garcia-Quismondo, G. Mena Marugan.
	``The Martin-Benito-Mena Marugan-Olmedo prescription for the Dapor-Liegener model of Loop Quantum Cosmology''.
	{\it Phys. Rev. D} {\bf 99}, 083505
	(2019)	
	
\bibitem{GM19b}
	A. Garcia-Quismondo, G. Mena Marugan.
	``Dapor-Liegener formalisam of loop quantum cosmology for Bianchi I spacetimes''.
	[arXiv:1911.09978] (2019)
	

\bibitem{APS06a}
	A. Ashtekar, T. Paw{\l}owski, P Singh.
	``Quantum Nature of the Big Bang''.
	{\it Phys. Rev. Let.} {\bf 96}, 141301
	(2006)

\bibitem{APS06c}
	A. Ashtekar, T. Paw\l owski, P. Singh.
	``Quantum Nature of the Big Bang: Improved dynamics''.
	{\it Phys. Rev. D} {\bf 74}, 084003
	(2006)

\bibitem{Boj08}
	M. Bojowald.
	``Absence of a Singularity in Loop Quantum Cosmology''.
	{\it Phys. Rev. Lett.} {\bf 86}, 5227
	(2001)
	
\bibitem{Warsaw1}	
	E. Alesci, M. Assanioussi, J. Lewandowski.
	``A curvature operator for LQG''.
	{\it Phys. Rev. D}, {\bf 89}, 124017
	(2014)
	
\bibitem{Warsaw2}
	M. Assanioussi, J. Lewandowski, I. M\"akinen.
	``New scalar constraint operator for loop quantum gravity''.
	{\it Phys. Rev. D} {\bf 92}, 044042
	(2015)

\bibitem{ST04}
	P Singh, A. Toporensky.
	``Big Crunch Avoidance in k = 1 Semi-Classical Loop Quantum Cosmology''.
	{\it Phys. Rev. D} {\bf 69}, 104008	
	(2004)

\bibitem{SKL06}
	L. Szulc, W. Kami\'nski, J Lewandowksi.
	``Closed FRW model in Loop Quantum Cosmology''.
	{\it Class. Quant. Grav.} {\bf 24}, 2621-2636
	(2007)
		
\bibitem{APSV07}
	A. Ashtekar, T. Paw\l owski, P. Singh, K. Vandersloot.
	``Loop quantum cosmology of k=1 FRW models''.
	{\it Phys. Rev. D} {\bf 75}, 024035
	(2007)
	
\bibitem{MHS09}
	J. Mielczarek, O. Hrycyna, M. Szydlowski.
	``Effective dynamics of the closed loop quantum cosmology''.
	{\it JCAP} 0911:014
	(2009)

\bibitem{TW1}
	T. Thiemann.
	``Gauge Field Theory Coherent States (GCS): I. General Properties".
	{\it Class. Quant. Grav.} {\bf 18} 2025-2064
	(2001)

\bibitem{TW2}
	T. Thiemann, O. Winkler.
	``Gauge Field Theory Coherent States (GCS): II. Peakedness Properties".
	{\it Class. Quant. Grav.} {\bf 18} 2561-2636
	(2001)

\bibitem{TW3}
	T. Thiemann, O. Winkler.
	``Gauge Field Theory Coherent States (GCS): III. Ehrenfest Theorems".
	{\it Class. Quant. Grav.} {\bf 18} 4629-4682
	(2001)
	
\bibitem{Tav08}
	V. Taveras.
	``LQC corrections to the Friedmann equations for a universe with a free scalar
field''.
	{\it Phys. Rev. D} {\bf 78}, 064072
	(2008)

\bibitem{BS06}
	M. Bojowald, A. Skirzewski.
	``Effective theory for the cosmological generation of structure''.
	{\it Rev. Math. Phys.} {\bf 18}, 713
	(2006)	
	
\bibitem{GP19}
	S. Gielen, A. Polaczek.
	``Generalised effective cosmology from group field theory''.
	 [arXiv:1912.06143]
	(2019)	
	
\bibitem{Thi00}
	T. Thiemann.
	``Quantum Spin Dynamics (QSD) : VII. Symplectic Structures and Continuum Lattice Formulations of Gauge Field Theories’’.
{\it Class. Quant. Grav.} {\bf 18}, 3293-3338
(2001)



\bibitem{DKL20}
	W. Kami\'nski, K. Liegener.
	``Symmetry restriction and its application to gravity''.
	[arXiv:2009.06311]
	(2020)
	
\bibitem{HL19}
	M. Han, H. Liu.
	``Effective Dynamics from Coherent State Path Integral of Full Loop Quantum Gravity''.
	[arXiv:1910.03763]
	(2019)
	
\bibitem{AL98}
	A. Ashtekar, J. Lewandowski.
	``Quantum Theory of Geometry II: Volume operators''.
	{\it Adv. Theor. Math. Phys.} 1, 388-429	
	(1998)
	

\bibitem{DLP19}
	A. Dapor, K. Liegener, T. Paw\l owksi.
	``Challenges in recovering a consistent cosmology from the effective dynamics of loop quantum gravity''.
	{\it Phys. Rev. D} {\bf 100}, 106016
	(2019)
	
\bibitem{Bod15}
	N. Bodendorfer.
	``An embedding of loop quantum cosmology in (b, v) variables into a full theory context''.
	{\it Class. Quantum Grav.} 33, 125014
	(2016)
	
\bibitem{HL19_Improv}
    M. Han, H. Liu.
    ``Improved ($\overline{\mu}$-Scheme) Effective Dynamics of Full Loop Quantum Gravity''.
    {\it [arXiv:1912.08668]}
    (2019)

\bibitem{ADL19}
	M. Assanioussi, A. Dapor, K. Liegener
	``Perspectives on the dynamics in loop effective black hole interior''.
	{\it Phys. Rev. D} {\bf 101}, 026002 
	(2020)
		

	}
\end{thebibliography}
\end{document}